# Strain Solitons in an Epitaxially Strained van der Waals-like Material

Jason T. Dong[1], Hadass S. Inbar[1], Connor P. Dempsey[2], Aaron N. Engel[1], Christopher J. Palmstrøm[1,2*]

[1]Materials Department, University of California, Santa Barbara, CA 93106

[2]Deparment of Electrical and Computer Engineering, University of California, Santa Barbara, CA 93106

Abstract

Strain solitons are quasi-dislocations that form in van der Waals materials to relieve the energy associated with lattice or rotational mismatch in the crystal. Novel and unusual electronic properties of strain solitons have been both predicted and observed. To date, strain solitons have only been observed in exfoliated crystals or mechanically strained bulk crystals. The lack of a scalable approach towards the generation of strain solitons poses a significant challenge in the study of and use of the properties of strain solitons. Here we report the formation of strain solitons with epitaxial growth of bismuth on an InSb (111)B substrate by molecular beam epitaxy. The morphology of the strain solitons for films of varying thickness is characterized with scanning tunneling microscopy and the local strain state is determined from the analysis of atomic resolution images. Bending in the solitons is attributed due to interactions with the interface, and large angle bending is associated with edge dislocations. Our results enable the scalable generation of strain solitons.

*cjpalm@ucsb.edu

Strain solitons are topological defects in van der Waals materials that act as quasi-dislocations to relax strain [1]. The strain solitons are local regions of the crystal where the atoms have rearranged to relieve elastic strain and the local symmetry is broken. Novel electronic properties such as local band gap modification [2], charge carrier confinement [3], flatbands [4-6], and topological edge modes [7,8] have been both predicted and observed to occur within the strain solitons. Strain solitons have been experimentally observed in mechanically strained bulk crystals [3], exfoliated graphene layers [1,4,7-10], hBN multilayers [11], and twisted transition metal dichalcogenides [2,12]. However, a wafer scale approach for the generation of strain solitons has yet to be demonstrated, which poses a significant challenge for both the investigation and the utilization of the properties of strain solitons. Here, to the best of the author's knowledge, we are the first to report the formation of strain solitons in bismuth thin films with heteroepitaxial growth on a III-V semiconductor substrate, which enables the scalable creation of strain solitons in a van der Waals-like material.

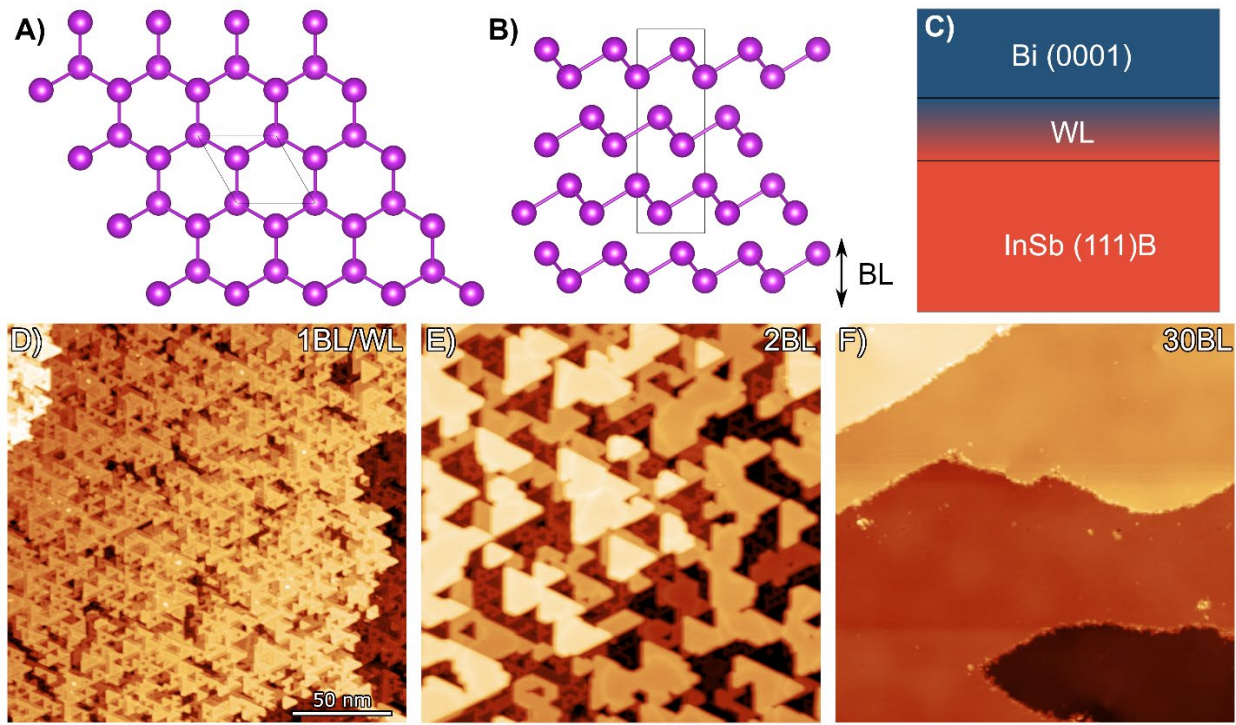

Figure 1: A) Top view of an individual bilayer (BL) of bismuth, unit cell of bismuth is outlined. B) Side view of a bismuth BL, with the unit cell outlined. C) Layer schematic, consisting of an InSb (111)B substrate, a wetting layer (WL), and the bismuth (0001) thin film. 200 nm x 200 nm STM images of: D) the bismuth WL on the InSb surface after deposition of 1 BL of bismuth ($I_T$ = 5pA, $V_B$ = 3V), E) a 2 BL thick film, with bismuth islands growing on the wetting layer ($I_T$ = 5pA, $V_B$ = 3V), and F) a 30 BL thick film, showing a complete film with 1 BL terraces present ($I_T$ = 5pA, $V_B$ = 3V).

Bismuth is a van der Waals-like material (R$\bar{3}$m), the crystal structure can be understood as being quasi-hexagonal with buckled layers (1 bilayer) of bismuth in a honeycomb lattice (Fig. 1A & B). In the bulk, bismuth is a semimetal with large spin-orbit coupling and has demonstrated evidence of a higher order topology [13]. In this study, strain solitons in bismuth are generated with heteroepitaxial growth of bismuth on InSb (111)B by molecular beam epitaxy. From the bulk lattice constants, the InSb (111)B substrate is expected to apply a 0.8% tensile strain to the bismuth thin film. The layer schematic of the material system is shown in Fig. 1C, where the layer structure comprises of the InSb (111)B substrate, a bismuth wetting layer, and the bismuth thin film. The first bilayer (BL) of bismuth deposited on InSb forms a wetting layer comprised of atoms arranged in a Sierpiński triangle like pattern [14] (Fig. 1D). Upon subsequent deposition of additional bismuth, islands of (0001) bismuth grow epitaxially on the wetting layer (Fig. 1E). While ultrathin bismuth thin films have been observed to grow in the black phosphorous crystal structure [15], the quasi-hexagonal bismuth phase is observed to grow in this study for all thicknesses, which likely forms due to the interactions with the substrate. The bismuth islands appear to be oriented in the same direction, which is consistent with the absence of rotational twinning. After additional deposition of bismuth, the islands of bismuth coalesce into a complete, epitaxial single crystal thin film (Fig. 1F). See supplementary information S1 for additional XRD of the epitaxial bismuth.

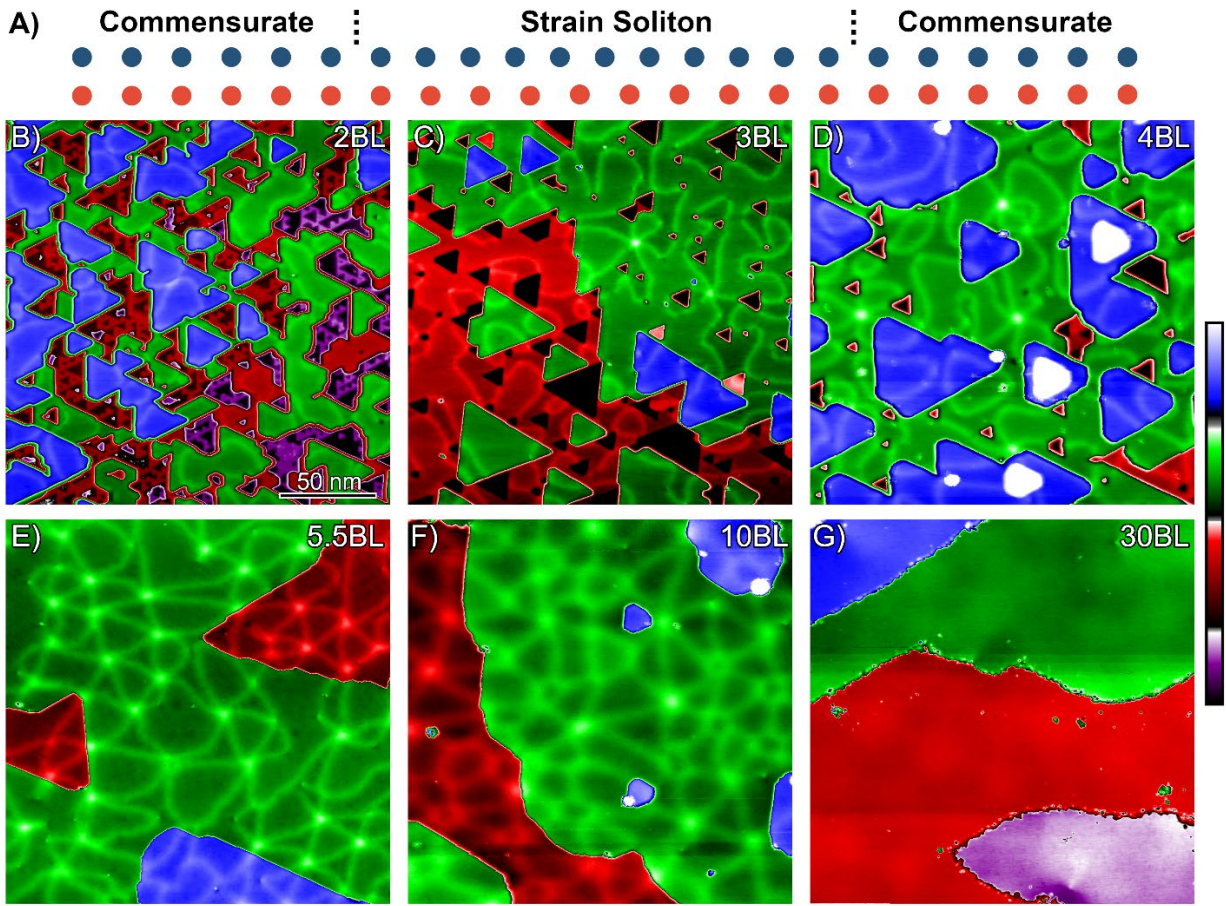

Figure 2: A) Schematic of a strain soliton, acting as the domain wall between commensurate regions of the crystal in the thin film (top layer) on the substrate (bottom layer). 200 x 200 nm STM images of B) 2BL thick film, C) 3 BL thick film, D) 4 BL thick film, E) 5.5 BL thick film, F) 10 BL thick film, and G) 30 BL thick film ($I_T$ = 5pA, $V_B$ = 3V).

Strain soliton formation (also termed replications) in the bismuth thin films is due to relaxation of the epitaxial tensile strain from the InSb substrate. The strain solitons are local regions of incommensurate bismuth that are relaxed, these solitons separate commensurate (strained) regions of bismuth (Fig. 2A). The relaxation of the bismuth is visible in the STM images of the bismuth thin films as local regions of brighter contrast, the brighter contrast is predominately an out of plane buckling of the thin film due to the soliton relaxation, with little electronic state contribution to the contrast (see supplementary information S2). In the 2 BL thick film (Fig. 2B), soliton lines can be observed to start and terminate at the edges of the bismuth islands. As the bismuth thin film becomes thicker the soliton morphology can be observed to evolve from predominantly soliton loops in the 3 BL thick film (Fig. 2C), to loops and nodes (Fig. 2D), and finally to a soliton network (Fig. 2E & F). The strain solitons propagate through the different bilayer steps of the bismuth thin film, indicating the strain solitons form at the interface between the bismuth thin film and the wetting layer and propagate throughout the thin film. Upon deposition of the 30 BL of bismuth, no evidence of strain solitons can be observed by STM and the film appears to have completely relaxed. See supplementary information of XRD supporting this relaxation by forming incommensurate and commensurate regions.

Analogous to characterization of dislocations, strain solitons can be characterized with the direction of their displacement vector and the direction of propagation [1]. Tensile strain solitons have a displacement vector perpendicular to the direction of propagation, while shear strain solitons have a displacement vector parallel to the direction of propagation. Tensile and shear solitons have different directions of propagation in the crystal. In materials comprised of honeycomb lattices, tensile strain solitons propagate along the zigzag or <$10\bar{1}0$> direction of the crystal while shear solitons propagate along the armchair or <$11\bar{2}0$> direction of the crystal. Under a biaxial tensile strain, it is expected that a regularly spaced network of tensile strain solitons[16,17]. However, irregularly spaced nodes in the soliton networks are observed in the bismuth. Additionally, the bending of the strain solitons is also visible in the STM images for all thickness in which strain solitons are visible. As a result of this bending, the strain solitons are of tensile, shear, and mixed type in the bismuth thin film.

To gain further insight into the bending mechanism of the strain solitons, we examine an atomic resolution STM image of a 5.5 BL thick film where both a large angle bend and a small angle bend in the strain solitons is observed (Fig. 3A). Fig. 3B and 3C are enlargement of the large angle and small angle bends to resolve the atoms, respectively. The large angle bend has a soliton bend angle of 150° with the strain soliton converting from a tensile strain soliton to a shear strain soliton, and a point of brighter contrast can be observed at the bend. This point of brighter contrast is an edge dislocation core. The out of plane edge dislocation has an extra half plane in-plane of the bismuth thin film. From the burgers circuit, the burgers vector is $a[01\bar{1}0]$, with $a$ being the in plane lattice constant of the bismuth crystal. The small angle bend of 30° and is also where the strain soliton converts from a tensile strain soliton to a shear strain soliton. However, at this small angle bend there is no edge dislocation present. The edge dislocations are observed at bends of greater than 30°. We propose that this correlation between the large angle bending of solitons and edge dislocation a locally anisotropic strain field favoring the bending of a strain soliton and the nucleation of an edge dislocation to bend the strain soliton and relieve the anisotropic strain, similar to what has been observed before in uniaxially strain $MoSe_2$ [3].

The bending and edge dislocation formation is indicative that there is inhomogeneous strain relaxation. The inhomogeneous strain is likely induced in the thin films, instead of the expected homogenous, biaxial tensile strain from heteroepitaxial growth. We attribute the local inhomogenous strain due to interactions between the bismuth and the Sierpiński structure of the wetting layer. Additionally, the change in strain soliton morphology from isolated loops to a soliton network, with the strain solitons becoming predominantly tensile strain solitons as the film thickness increases, is likely due to local interactions with the wetting layer dominate in the thinner films and induce a locally inohomgenous strain which results in the strain soliton morphology to favor bending and loop formation. However, as the film thickness increases these interactions influence is lessened and the biaxial strain from the substrate favors the formation of a strain soliton network of tensile strain solitons.

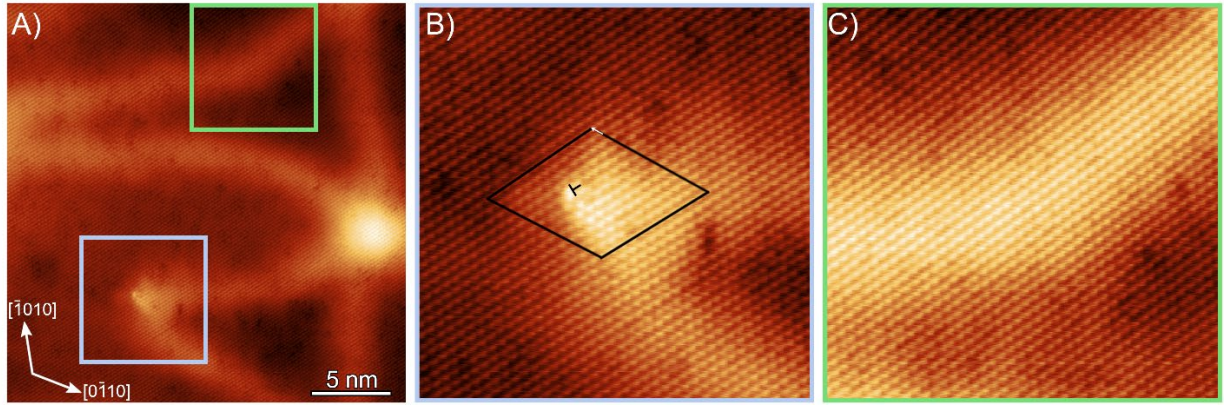

Figure 3: A) Atomic resolution image of 5.5 BL thick film ($I_T$ = 200pA, $V_B$ = 0.25V). B) STM image of a large angle soliton bend, with an edge dislocation present ($I_T$ = 200pA, $V_B$ = 0.25V). From, the burgers circuit the burgers vector is $a[01\bar{1}0]$. C) STM image of a small angle (30º) bend, no edge dislocations are present at this bend ($I_T$ = 200pA, $V_B$ = 0.25V).

Having now studied the bending of the strain solitons, the strain state of the strain solitons in the bismuth thin film is investigated with strain maps from atomic resolution STM images. Fig. 4A shows an atomic resolution image of a 5.5 BL thick film, where a soliton node is observed along with a tensile strain soliton and a shear strain soliton. Fig. 4B is an enlargement of the image around the nodes where the atoms of the bismuth thin film are clearly visible. In the atomic resolution image, only the atoms in the top layer of the bismuth bilayer are imaged, generating the observed hexagonal lattice of atoms. The in-plane strain from the atomic resolution image is determined from the atomic displacements measured with the Lawler-Fujita algorithm [18], which is a method of determining atomic displacements similar to geometric phase analysis [19]. See the supplementary information S3 for additional details on the Lawler-Fujita algorithm. The Bragg peaks used for the analysis are circled in the Fourier transform of the atomic resolution image shown in Fig. 4C. The in-plane strain maps of the strain solitons and node are shown in Fig. 4D-F. From the maps it can be observed that a relaxation of 2.9% $\varepsilon_{xx}$ tensile strain and 1.3% $\varepsilon_{xy}$ shear strain is concentrated at the node. Additionally, the tensile strain soliton has the expected relaxation in the direction perpendicular to the direction of propagation, with a maximum relaxation of 1.6% $\varepsilon_{xx}$. Finally, it can be observed that within a shear strain soliton there is 1.4% shear strain present and negligible normal strain present. However, it can also be noted that there is tensile strain relaxation in the direction parallel to the direction of propagation, $\varepsilon_{yy}$, present in the vicinity of the shear soliton. The observed tensile relaxation is consistently larger than 0.8%, which is the expected tensile strain induced from the InSb substrate. This larger than expected relaxation is likely due to the reduction of lattice constant of bismuth in the ultrathin limit [20,21], resulting in larger strains being induced in the thin film when it is strained to InSb for ultrathin films. We observe evidence of the reduction in the lattice constant of relaxed bismuth in our thin films from our reciprocal space maps (see supplementary material S1).

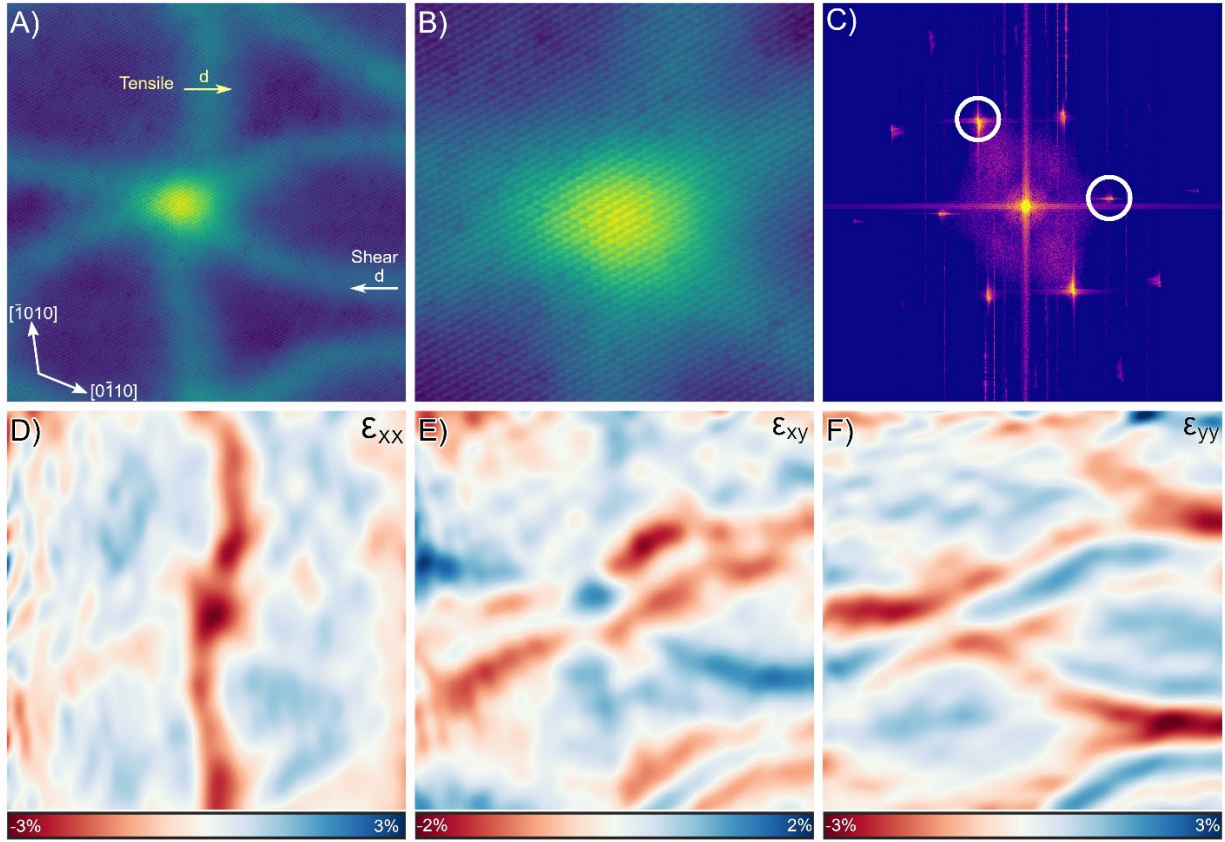

Figure 4: A) 43.8 x 43.8 nm atomic resolution image of a soliton node in a 5.5 BL thick film, both a tensile and a shear strain soliton are present with their corresponding displacement vectors indicated. ($I_T$ = 200pA, $V_B$ = 0.25V). B) Enlargement of the region near the node, atoms are visible. C) FFT of the atomic resolution image, the Bragg peaks used for the strain analysis are circled. Strain maps of the image: D) $\varepsilon_{xx}$, E) $\varepsilon_{xy}$, and F) $\varepsilon_{yy}$.

In summary, we have demonstrated the formation of strain solitons in a van der Waals-like material with heteroepitaxial growth. The epitaxial strain enables the formation of strain solitons. The bending of strain solitons is observed, with large angle bends accompanied by edge dislocations. A change in strain soliton morphology from loops to a network is observed as the film thickness increases. Normal and shear strain is concentrated in the nodes of the strain solitons. These results indicate that the wetting layer induces an inhomogeneous strain state in the bismuth thin film. Our results demonstrate the importance of interface interactions on the morphology of the strain solitons and enable the wafer scale generation of solitons and soliton networks.

**Methods**

Bi thin films where grown by MBE on unintentionally doped InSb (111)B (Wafer Technology Ltd.) substrates. The substrate surface was cleaned with atomic hydrogen cleaning, see [22] for additional details on the surface preparation, which resulted in the (3x3) surface reconstruction. Thin films were nucleated at 14 °C followed by low temperature annealing at 80-120 °C for several hours to improve the film morphology. All thickness reported in this study are nominal thickness based on the calibration of the growth rate from Rutherford backscattering spectrometry.

*In-vacuo* STM measurements were performed with an Omicron LT STM at 78 K. Mechanically cut PtIr tips prepared with field emission on Au foil were used for the STM measurements.

**Supporting Information**

Details of the strain mapping, additional STM images, and XRD of the thin films.

**Acknowledgements**

This work was supported by University of California Multiple Campus Award No. 00023195. We acknowledge the use of shared facilities of the NSF Materials Research Science and Engineering Center (MRSEC) at the University of California Santa Barbara (Grant No. DMR 2308708).

**References**

[1] Alden, J. S.; Tsen, A. W.; Huang, P. Y.; Hovden, R.; Brown, L.; Park, J.; Muller, D. A.; McEuen, P. L.; *PNAS* **2013**, *110*, 11256-11260. DOI: 10.1073/pnas.1309394110
[2] Rosenberger, M. R.; Chuang, H.-J.; Phillips, M.; Oleshko, V. P.; McCreary, K. M.; Sivaram, S. V.; Hellberg, C. S.; Jonker, B. T.; *ACS Nano* **2020**, *14*, 4550-4558. DOI: 10.1021/acsnano.0c00088
[3] Edelberg, D.; Kumar, H.; Shenoy, V.; Ochoa, H.; Pasupathy, A. N.; *Nature Physics*, **2020**, *16,* 1097-1102. DOI: 10.1038/s41567-020-0953-2
[4] Kazmierczak, N. P.; Van Winkle, M.; Ophus, C.; Bustillo, K. C.; Carr, S.; Brown, H. G.; Ciston, J.; Taniguchi, T.; Watanabe, K.; Bediako, D. K.; *Nature Materials* **2021**, *20*, 956-963. DOI: /10.1038/s41563-021-00973-w
[5] Naik, M. H.; Jain, M.; *PRL* **2018**, *121*, 266401. DOI: 10.1103/PhysRevLett.121.266401
[6] Kang, D.; Zuo, Z.-W.; Wang, Z.; Ju, W.; *RSC Adv.* **2021**, *11*, 24366-24373. DOI: 10.1039/d1ra02139g
[7] Ju, L.; Shi, Z.; Nair, N.; Lv, Y.; Jin, C.; Velasco Jr, J.; Ojeda-Aristizabal, C.; Bechtel, H. A.; Martin, M. C.; Zettl, A.; Analytis, J.; Wang, F.; *Nature* **2015**, *520*, 650-655. DOI: 10.1038/nature14364
[8] Yoo, H.; Engelke, R.; Carr, S.; Fang, S.; Zhang, K.; Cazeaux, P.; Sung, S. H.; Hovden, R.; Tsen, A. W.; Taniguchi, T.; Watanabe, K.; Yi, G.-C.; Kim, M.; Luskin, M.; Tadmor, E. B.; Kaxiras, E.; Kim, P.; *Nature Materials* **2019**, *18*, 448-453. DOI: 10.1038/s41563-019-0346-z
[9] Jiang, L.; Shi, Z.; Zeng, B. Wang, S.; Kang, J.-H.; Joshi, T.; Jin, C.; Ju, L.; Kim, J.; Lyu, T.; Shen, Y.-R.; Crommie, M.; Gao, H.-J.; Wang, F.; *Nature Materials* **2016**, *15*, 840-844. DOI: 10.1038/NMAT4653
[10] Jiang, L.; Wang, S.; Shi, Z.; Jin, C.; Utama, M. I. B.; Zhao, S.; Shen, Y.-R.; Gao, H.-J.; Zhang, G.; Wang, F.; *Nature Nanotechnology* **2018**, *13*, 204-208. DOI: 10.1038/s41565-017-0042-6

[11] Ni, G. X.; Wang, H.; Jiang, B.-Y.; Chen, L. X.; Du, Y.; Sun, Z. Y.; Goldflam, M. D.; Frenzel, A. J.; Xie, X. M.; Fogler, M. M.; Basov, D. N.; *Nature Communications* **2019**, *10*, 4360. DOI: 10.1038/s41467-019-12327-x

[12] Tilak, N.; Li, G.; Taniguchi, T.; Watanabe, K. Andrei, E. Y.; *Nano Letters* **2023**, *23,* 73-81. DOI: 10.1021/acs.nanolett.2c03676

[13] Schindler, F.; Wang, Z.; Vergniory, M. G.; Cook, A. M.; Murani, A.; Sengupta, S.; Kasumov, A. Y.; Deblock, R.; Jeon, S.; Drozdov, I.; Bouchiat, H.; Guéron, S.; Yazdani, A.; Bernevig, B. A.; Neupert, T.; *Nature Physics* **2018**, *14*, 918-624. DOI: 10.1038/s41567-018-0224-7

[14] Liu, C.; Zhou, Y.; Wang, G.; Yin, Y.; Li, C.; Huang, H.; Guan, D.; Li, Y.; Wang, S.; Zheng, H.; Liu, C.; Han, Y.; Evans, J. W.; Liu, F.; Jia, J.; *PRL* **2021**, *126*, 176102. DOI: 10.1103/PhysRevLett.126.176102

[15] Nagao, T.; Sadowski, J. T.; Saito, M.; Yaginuma, S.; Fujikawa, Y.; Kogure, T.; Ohno, T.; Hasegawa, Y.; Hasegawa, S.; Sakurai, T.; *PRL* **2004**, 93, 105501. DOI: 10.1103/PhysRevLett.93.105501

[16] Lebedeva, I. V.; Popov, A. M.; *PRB* **2019**, *99*, 195448. DOI: 10.1103/PhysRevB.99.195448

[17] Levedeva, I. V.; Popov, A. M.; *J. Phys. Chem. C.* **2020**, *124*, 2120-2130. DOI: 10.1021/acs.jpcc.9b08306

[18] Lawler, M. J.; Fujita, K.; Lee, J.; Schmidt, A. R.; Kohsaka, Y.; Kim, C. K.; Eisaki, H.; Uchida, S.; Davis, J. C.; Sethna, J. P.; Kim, E.-A.; *Nature* **2010**, *466*, 347-351. DOI: 10.1038/nature09169

[19] Hÿtch, M. J.; Snoeck, E.; Kilaas, R.; *Ultramicroscopy* **1998**, *74*, 131-146. DOI: 10.1016/S0304-3991(98)00035-7

[20] Lisgarten, N. D.; Peppiatt, S. J.; Sambles, J. R.; *J. Phys. C: Solid State Phys.* **1974**, *7*, 2263-2268. DOI: 10.1088/0022-3719/7/13/006

[21] Cantele, G.; Ninno, D.; *Physical Review Materials* **2017**, *1*, 014002. DOI: 10.1103/PhysRevMaterials.1.014002

[22] Dong, J. T.; Inbar, H. S.; Pendharkar, M.; van Schijndel, T. A. J.; Young, E. C.; Dempsey, C. P.; Palmstrøm, C. J.; *JVST B* **2023**, *41*, 032808. DOI: 10.1116/6.0002606